\documentclass[aps,prl,twocolumn,superscriptaddress,showpacs]{revtex4-1}
\usepackage{epsfig,amssymb,amsmath,theorem,graphicx,subfigure}
\newcommand{\ud}{\mathrm{d}}

\begin{document}

\title{Force localization in contracting cell layers}

\author{Carina M.  Edwards}
\affiliation{University of Heidelberg, Bioquant, Im Neuenheimer Feld 267, 69120 Heidelberg, Germany}
\author{Ulrich S. Schwarz}
\email[]{Ulrich.Schwarz@bioquant.uni-heidelberg.de}
\affiliation{University of Heidelberg, Bioquant, Im Neuenheimer Feld 267, 69120 Heidelberg, Germany}
\affiliation{University of Heidelberg, Institute for Theoretical Physics, Philosophenweg 19, 69120 Heidelberg, Germany}

\date{\today}

\begin{abstract}
Epithelial cell layers on soft elastic substrates or pillar arrays are
commonly used as model systems for investigating the role of force in
tissue growth, maintenance and repair. Here we show analytically that
the experimentally observed localization of traction forces to the
periphery of the cell layers does not necessarily imply increased
local cell activity, but follows naturally from the elastic problem of
a finite-sized contractile layer coupled to an elastic foundation. For
homogeneous contractility, the force localization is determined by one
dimensionless parameter interpolating between linear and exponential
force profiles for the extreme cases of very soft and very stiff
substrates, respectively.  If contractility is sufficiently increased
at the periphery, outward directed displacements can occur at
intermediate positions, although the edge itself still retracts. We
also show that anisotropic extracellular stiffness leads to force
localization in the stiffer direction, as observed experimentally.
\end{abstract}

% 87.10.+e general theory biological physics
% 87.16.Ln cytoskeleton
% 87.17.Rt cell adhesion and cell mechanics


\maketitle

Actively generated forces have emerged as a central element in the way
tissue cells interact with their environment, for example during
development and tissue maintenance \cite{WozniakChen2009}. It
is becoming increasingly clear that cells use physical force to probe
the mechanical properties of their environments and to create a
structurally coherent state within the cell and
tissue. Experimentally, however, it is challenging to measure cellular
forces in a physiological context. Therefore their effect is usually
assessed in an indirect manner, for example by mechanical relaxation
after laser cutting \cite{KiehartEdwards2003}.

In order to directly measure the forces that cells exert on their
environments, two complementary techniques have been established over
the last decade. Embedding marker beads into soft elastic gels enables
the measurement of the displacement field generated by cell traction
forces applied to the surface of the gel, which can then be estimated
using elasticity theory \cite{DemboWang1}.
Alternatively, microfabrication techniques are used to create an array
of elastomeric pillars, which for small displacements each act as a
linear spring \cite{TanTienPirone}. 

In order to address the role of forces for larger cell clusters, both
soft elastic substrates and pillar assays have been used for confluent
layers of epithelial cells. Placing such cells on a pillar array, it
was found that cellular traction forces are localized to the edge of
the cell layer \cite{duRoureSaezBuguin,SaezAnonGhibaudo}. A
qualitatively similar effect has also been observed for cell sheets
migrating over soft elastic substrates \cite{TrepatWasserman2009},
although in this case the length scales over which the stresses
decayed were greater.  It is tempting to assume that force
localization to the edge reflects increased mechanical activity of the
cells at the edge, for example of putative leader cells
\cite{KhalilFriedl}. However, a more detailed analysis requires a mechanical
analysis of the cell sheet coupled to the elastic foundation.

In this Letter, we analytically solve the problem of a contracting
cell layer coupled to a layer of springs. For homogeneous contraction,
we show that the problem is determined by one dimensionless parameter
which interpolates between the extreme cases of soft pillars with a
linear force profile and stiff pillars with an exponential force
profile.  We also show how our results change for increased contractile
activity at the rim and for substrates with anisotropic stiffness.

\begin{figure}
\includegraphics[width=0.48\textwidth]{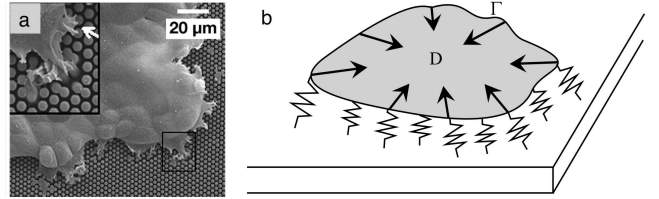}
\caption{(a) Scanning electron micrographs of an epithelial cell
  layer on an array of microfabricated elastomeric pillars (image reproduced from
  \cite{duRoureSaezBuguin}.) (b) A
  schematic representation of the pillar assay.  The cell layer contracts and is
  resisted by a distribution of linear springs on the surface.}
\label{fig:Ladoux1}
\end{figure}

\textit{Elastic model.} Microstructured surface assays consist of a
regular array of flexible elastomeric pillars with defined surface
chemistry, see Fig.~\ref{fig:Ladoux1}. For small displacements and
slender pillars, the relation between the traction applied to the
pillars $\mathbf{T}$ and the displacement $\mathbf{u}$ follows from
linear elasticity theory \cite{LandauLifschitz},
\begin{eqnarray} \label{springconstant}
\mathbf{T}&=&k\mathbf{u}\,=\,\left(\frac{3 \pi}{4} E_\mathrm{p} \frac{r^4}{L^3}\right) \mathbf{u},
\end{eqnarray}
where $r$ and $L$ are the radius and height of the pillar,
respectively, and $E_\mathrm{p}$ is the Young's modulus.  The pillars
thus behave as simple linear springs with effective spring constant
$k$.

In the assays of interest the substrate is usually coated with
adhesion-promoting molecules so that the cells spread thinly and form
a confluent layer. We thus assume the layer behaves as a
continuous elastic solid. In addition, as the typical lateral
extension of the sheet is much larger than the layer thickness $h$, we
have the conditions for plane-stress in the layer and can use a
two-dimensional model for the cells. The force balance then reads
\begin{eqnarray}
\mathrm{div} \cdot \mathbf{F} -k N \mathbf{u}&=&\mathbf{0}, 
\label{eq:Equilibrium}
\end{eqnarray}
where $\mathbf{F}$ is the two-dimensional in-plane stress tensor
obtained by averaging the layer stress over the
thickness, and $N$ is the number density of pillars.  For a continuous
elastic substrate, the restorative force applied to the cell layer
would not take this simple form. Moreover more detailed assumptions
would be required to define the elastic problem of the two coupled
elastic layers.  Nevertheless, the linear traction force term from
Eq.~\ref{eq:Equilibrium} can be also considered as a first order
approximation for soft elastic substrates \cite{MurrayBook2,EdwardsChapman}.

The cell layer is mechanically active, with cells simultaneously
repositioning and contracting. Here we consider only the final
configuration of a contractile cell layer.  Then the problem of
modelling layer contraction is similar to that of thermoelasticity,
where temperature changes generate expansion or contraction in the
material \cite{MurrayBook2}. In thermoelasticity one typically
introduces an expansion term proportional to the target volume change
into the constitutive relation \cite{LandauLifschitz}.  In analogy we
take
\begin{equation}
\label{eq:Constitutive}
F_{ij} = \frac{h E_\mathrm{c}}{1+\nu}\left( e_{ij}+\frac{\nu}{1-\nu}e_{kk}\delta_{ij} \right)
-\frac{h E_\mathrm{c}}{2(1-\nu)} P \delta_{ij} .
\end{equation}
The first term is the usual plane-stress linear elastic constitutive
relation with $E_\mathrm{c}$ the cellular Young's modulus, $\nu$ the
cellular Poisson's ratio, and $e_{ij}$ the strain tensor.  The second
term in Eq.~(\ref{eq:Constitutive}) captures cellular contractility,
where we have assumed that the contraction is isotropic. If the layer
is under no stress, $F_{ij}=0$ and thus $P=e_{ii}$ (summation convention applies), and so $P$ is
the local target volume change of a material element in the cell layer.

At the layer boundary no stress is applied, giving the boundary
condition $\mathbf{F} \cdot \mathbf{n} = \mathbf{0}$ on
$\Gamma$. Integrating Eq.~(\ref{eq:Equilibrium}) over the cell area
$D$ and applying the zero-stress boundary condition gives $\int_{D} k
N \mathbf{u}=\mathbf{0}$.  Thus momentum is not generated in the
system, as expected for a closed system.

\textit{Solution for a contracting stripe.} It is instructive to first
solve our elastic problem in one dimension, although experimentally
even very anisotropic cell layers always contract in both lateral
directions. We consider a rectangular sheet with $e_{yy}=0$,
such that the layer contracts only in the $x$-direction over
the stripe width $l_0$.  Substituting Eq.~(\ref{eq:Constitutive}) into Eq.~(\ref{eq:Equilibrium}) we find that $u(x)$ satisfies
\begin{align}
\frac{\ud^2 u}{\ud x^2}-  \frac{1}{l^2} u&\,=\,\frac{1}{2}(1+\nu)\frac{\ud P}{\ud x} \label{eq:ODE1}
\end{align}
with the zero stress boundary condition $\ud u / \ud x = (1+\nu)P / 2$
at $x=\pm l_0$ and with $u=0$ at $x=0$. Here we have introduced a new
length scale, the localization length
\begin{equation}
l = \sqrt{\frac{h E_\mathrm{c}}{k N (1-\nu^2)}} = \sqrt{\frac{4 h E_\mathrm{c} L^3}{3\pi (1-\nu^2) E_\mathrm{p} N r^4}},
\end{equation}
with the second expression following from Eq.~(\ref{springconstant}). 

\begin{figure}
\includegraphics[width=0.5\textwidth]{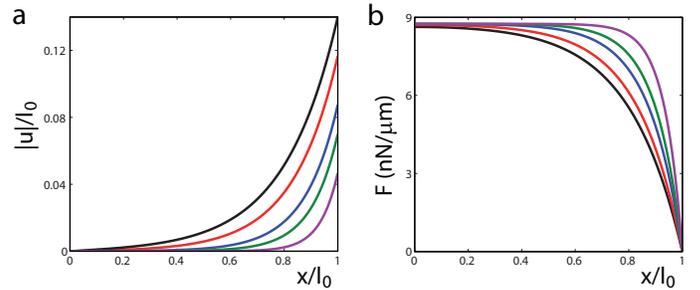}
\caption{(Color online) Plot of (a) displacement $|u|$ and (b) stress $F$ within the cell layer 
for the stripe geometry with $P_0=0.7$, 
$E_\mathrm{c}=10 \, \mathrm{kPa}$, $h=1\, \mu \mathrm{m}$, $\nu=0.45$ and $\gamma=$5, 6, 8, 10, 15
(from top to bottom in (a) and from bottom to top in (b)).}
\label{fig:LinearSolution}
\end{figure}

Assuming that the cell layer contracts uniformly, $P$ is constant.
For the stripe case, it is convenient to set $P=-2 P_0/(1+\nu)$ where
$P_0$ is then the target contraction in the $x$-direction. We then find that Eq.~(\ref{eq:ODE1}) is solved by
\begin{eqnarray} \label{eq:sol1}
\frac{u}{l_0}&=&-P_0  \frac{\mathrm{sinh} (\gamma x / l_0) } {\gamma  \mathrm{cosh} \gamma} 
\end{eqnarray}
where we have defined the dimensionless localization parameter $\gamma = l_0/l$, the
ratio of layer dimension to the localization length $l$.  The internal
stress $F(x)$ can be obtained by substituting Eq.~(\ref{eq:sol1}) into
the constitutive relation Eq.~(\ref{eq:Constitutive}), see SI \cite{SI}.

In Fig.~\ref{fig:LinearSolution} the displacement $u$ and internal stress
$F$ are plotted for different values of $\gamma$.  We see that
a constant cellular contraction leads to non-constant pillar
displacement, with greatest deflections at the edge.  The
non-dimensional parameter $\gamma$ controls the profile of the
solution and the rate of decrease of observed tractions away from the
layer edge. When $\gamma$ is large, the pillars dominate the system
and restrict the displacements to a small region near the rim. When
$\gamma$ is small, the substrate resists less and the tractions decay
slower. Note that both displacement and stress scale linearly with cell
contraction $P_0$.

\textit{Solution for a contracting disc.} We now consider the
two-dimensional problem of a circular cell layer with isotropic
contraction, which is directly comparable to experiments with
finite-sized and isotropically contracting cell layers. From symmetry
arguments $\mathbf{u}=u(r)\mathbf{e}_r$, where $r$ is the radial
distance from the centre of the plate.  The radially symmetric
equivalent to Eq.~(\ref{eq:ODE1}) reads
\begin{eqnarray}
r^2\frac{\ud^2 u}{\ud r^2}+r\frac{\ud u}{\ud r}-\left(1+\frac{r^2}{l^2}\right) u&=&\frac{1}{2}(1+\nu)r^2\frac{\ud P}{\ud r}.
\label{eq:circular}
\end{eqnarray} 
The boundary conditions are now $u = 0$ at $r= 0$, while at $r= r_0$ we have
$P = 2 ( \ud u / \ud r + u / r) / (1+\nu)$.

Considering again the case of constant cellular contraction, $P=-P_0$, 
the solution of Eq.~(\ref{eq:circular}) with the specified boundary conditions
can be calculated in terms of modified Bessel functions:
\begin{eqnarray}
\frac{u}{r_0}&=&-\frac{1}{2 \gamma}P_0(1+\nu) A(\gamma)I_1(\gamma r/r_0), \label{eq:circular11}\\
A(\gamma)&=&\left(I_0(\gamma) +\frac{\nu-1}{\gamma}I_1(\gamma)  \right)^{-1}. \label{eq:circular2}
\end{eqnarray}
Again $\gamma=r_0/l$ is the ratio of sample dimension and localization
length, i.e. the decay length scale for the traction forces $l$.  From
Eq.~(\ref{eq:circular11}) the radial stresses $F_{rr}$ and hoop
stresses $F_{\theta \theta}$ in the layer can be calculated, see SI \cite{SI}, which
show a similar behaviour as in the one-dimensional case from
Fig.~\ref{fig:LinearSolution}(b), except that the hoop stress does not
vanish at the boundary. The plot for displacement essentially looks
like the one for the stripe case, Fig.~\ref{fig:LinearSolution}(a), see SI \cite{SI}.
Our results are consistent with the simulations in
\cite{SpatzSimulate}, where a uniformly contracting network of actin
fibres on a discrete pillar array was also observed to exhibit
localisation of force to the edges of the network.

Regarding the comparison with experiments, typical parameter values
are $N = 0.25 \,\mu \mathrm{m}^{-2} $ for the pillar density, $r_0=50
\, \mu \mathrm{m}$ for layer dimension, $h=1\, \mu \mathrm{m}$ for
layer thickness, $E_\mathrm{c}=10 \, \mathrm{kPa}$ for cellular
Young's modulus and $\nu=0.45$ for Poisson's ratio. With pillar
stiffness ranging over 1 - 200 $\mathrm{nN} \, \mu \mathrm{m}^{-1}$
\cite{SaezAnonGhibaudo}, this gives a typical range for the
dimensionless parameter $\gamma\sim 7-100$. The clear decay of the
traction profile reported in \cite{SaezAnonGhibaudo} agrees nicely
with our results. A more spread-out traction profile has been reported
in \cite{TrepatWasserman2009}, which may be attributable to a low
value of $\gamma$. However, as this study focused on the dynamics in a
migrating sheet, direct comparison is difficult in this case.

To further examine the role of $\gamma$ in determining the profile of
the solution, we consider the limiting cases of $\gamma$ small and
large.  When $\gamma$ is small, the pillars are very weak and the
layer will be able to contract by a large amount. When $\gamma \ll 1$, we
obtain to lowest order a linear profile $u \sim -P_0r/2$, as one can
infer either directly from the differential equation or from the
Bessel function. Alternatively, in the case $\gamma \gg 1$, when the
springs dominate the behaviour of the sheet, the displacement
localizes strongly to the edge and we get an exponential profile,
\begin{eqnarray}
u&\sim&-\frac{P_0r_0(1+\nu)}{2\gamma}e^{-\gamma (1-r/r_0)}.
\end{eqnarray}
The exact profile of the solution is thus clearly controlled by the
non-dimensional parameter, $\gamma$, which quantifies the relative
strengths of pillars versus cells. 

\textit{Effect of a contractile rim.} Up to now, we have assumed that
the layer undergoes a uniform contraction. We now apply our model for
the contractile cell layer to two further situations of large
practical interest, namely non-homogeneous contraction in the layer
and anisotropic layer stiffness. We first consider what the
displacements would look like if the layer had a more mechanically
active rim.  We consider that the contraction jumps up
at the rim:
\begin{equation}
  P = \left\{ 
  \begin{array}{l l}
    -P_0 & \quad \text{$x < x_1$ or $r \leq r_1$ }\\
    -P_1 & \quad \text{$x_1 < x <l_0$ or $r_1<r<r_0$}\\
  \end{array} \right.,
  \label{eq:StepP}
\end{equation}
for both the contracting stripe and disc.  Eqs.~(\ref{eq:ODE1}) and
(\ref{eq:circular}) are unchanged, as are the boundary conditions
applied at both the origin ($u=0$) and edge of the layer (zero normal
stress).  At the interface between the two contractile regions
(i.e. at $x=x_1$ and $r=r_1$, for the stripe and disc case,
respectively), however, we impose the additional conditions that the
displacement $u$ and stresses $F$ (for stripes) and $F_{rr}$ (for
discs) are continuous across the interface. With these additional
boundary conditions it is possible to obtain analytical solutions for
the displacement $u$, see SI  \cite{SI}.

\begin{figure}
\includegraphics[width=0.5\textwidth]{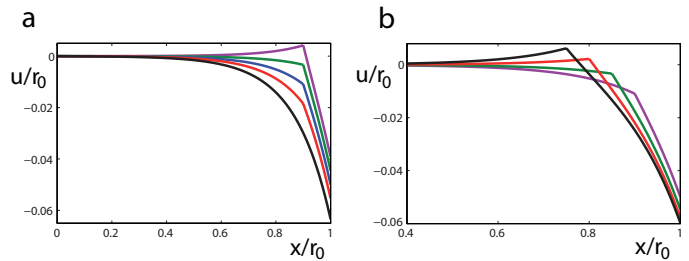}
\caption{(Color online) Plot of displacement $u$ in a disc with contractile rim for
(a) $P_0/P_1=$0.1, 0.3, 0.5, 0.7, 1  with $r_1/r_0=0.9$ (from top to bottom), (b) $r_1/r_0=$0.75, 0.8, 0.85, 0.9, with $P_0/P_1=0.5$ (from top to bottom). 
In both plots $\gamma=8$, $P_1=0.7$, $\nu=0.45$.}
\label{fig:StepP}
\end{figure}

For the disc geometry, the resulting displacement is plotted in
Fig.~\ref{fig:StepP}(a). Note that the plot of the displacement $u$
has a kink at the point of transition between the two contractilities,
and that as the inner region becomes weaker and $P_0/P_1$ decreases,
this feature is accentuated.  Equally, as $P_0/P_1\to 1$, this feature
disappears and we recover the solution for constant contractility. The
displacements for the contracting stripe solution are qualitatively
the same.

For a sufficiently inactive inner region, there is the possibility
that contraction of the outer rim dominates to such an extent that it
can pull the inner region towards it, resulting in positive
displacements at intermediate positions, although the layer as a whole
retracts its edge. This effect depends sensitively on the spatial
extent of the contracile rim. For the stripe case with
$x_1/x_0=1-\epsilon$ and $\epsilon \ll 1$, positive displacements are
possible only when $P_0 \sim \epsilon^2 \gamma^2 P_1/2$, see SI
\cite{SI}.  This dependence on the extent of the more contractile
region is also seen in Fig.~\ref{fig:StepP}(b). In summary, the
existence of kinks and positive displacements in experiments would be
a clear signature of increased contractility at the rim.

\textit{Solution for anisotropic pillars.} In a natural setting,
cells are often confronted with highly anisotropic environments, which can
have a strong effect on their response. This important effect can be
investigated experimentally by using pillars with elliptical
cross-sections \cite{TanTienPirone,LadouxOvalPillars,SaezAnonGhibaudo}.  Then the
apparent stiffness of the substrate depends on the direction,
\begin{equation}
k(\theta)=K\left(1 + \epsilon \cos^2 \theta \right),
\label{eq:ellipticalk}
\end{equation}
where $\theta$ is defined to be the angle of the pillar deflection
with respect to the major axis $a$.  The minor axis is given by the
relation $a^2=b^2(1+\epsilon)$. The spring constant is now $K=3 \pi
E_\mathrm{p} b^3 a/4 L^3$ and so we redefine $\gamma^2=r_0^2 K N (1-\nu^2)/h E_\mathrm{c}$.  Using
these arrays it is observed in
\cite{LadouxOvalPillars,SaezAnonGhibaudo} that tissues elongate along
the major axes of the pillars, with cellular force being concentrated
at the pointed ends of the layers.

\begin{figure}
\includegraphics[width=0.5\textwidth]{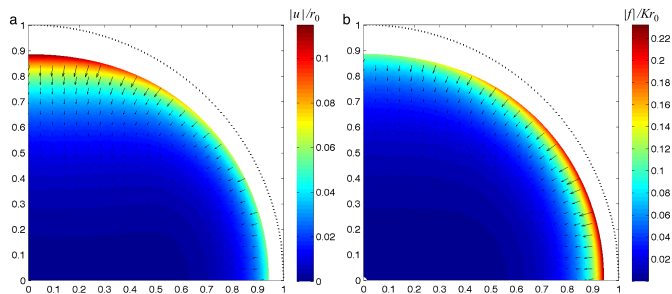}
\caption{(Color online) Plot of (a) displacement $|u|$, arrows
  indicate displacement directions and magnitude, (b) traction force $f/K r_0 $ exerted on pillars.  In both plots
  $\gamma=5$, $P_0=0.8$ and $\nu=0.45$, $\epsilon=3$ (major axis in
  $x$-direction), the dotted black line indicates the initial shape of
  the layer.}
\label{fig:FellipBoth}
\end{figure}

Combining Eqs.~(\ref{eq:Equilibrium}) and (\ref{eq:Constitutive}) with
$k$ given by Eq.~(\ref{eq:ellipticalk}), and with the boundary
conditions as before, the problem is now fully two-dimensional.  We
consider again the paradigm problem of an initially circular layer,
contracting isotropically, and solve the system numerically using
finite element methods for pillars with $a/b=2$ (i.e. $\epsilon=3$)
and the major axis in $x$-direction (MATLAB PDE toolbox, The
Mathworks, Natick, MA). For isotropic pillar elasticity, our numerical
solutions agree perfectly with the analytical results given above. The
numerical results for the anisotropic case are plotted in
Fig.~\ref{fig:FellipBoth}. Due to the symmetry of the problem, it is
sufficient to solve the problem on a quadrant, imposing zero $x,y$
displacement on $y=0, \, x=0$, respectively.  We see that an initially
circular layer on the anisotropic surface can be induced by uniform
isotropic contraction into a shape that is elongated in the direction
of larger stiffness ($x$-direction).  We also see that the traction
forces $f=k(\theta)\mathbf{u}$ are concentrated at the pointed ends of
the layer. We do not expect this effect to be sufficient to account
for the shape of the elongated tissue islands reported in
\cite{LadouxOvalPillars,SaezAnonGhibaudo}, but it is important to note
that an anisotropy in tissue shape and force profile can be induced
purely through the substrate properties.  Indeed this result of
uniform cellular tension generating anisotropic internal stresses
could be used by cellular assemblies to sense the direction of maximal
stiffness and to polarize accordingly.

\section{Acknowledgments}

CME was supported by a postdoctoral fellowship from the Center for
Modelling and Simulation in the Biosciences (BIOMS) at Heidelberg. USS
is a member of the Heidelberg cluster of excellence CellNetworks.

%\bibliography{refsPillars}

\begin{thebibliography}{14}%
\makeatletter
\providecommand \@ifxundefined [1]{%
 \@ifx{#1\undefined}
}%
\providecommand \@ifnum [1]{%
 \ifnum #1\expandafter \@firstoftwo
 \else \expandafter \@secondoftwo
 \fi
}%
\providecommand \@ifx [1]{%
 \ifx #1\expandafter \@firstoftwo
 \else \expandafter \@secondoftwo
 \fi
}%
\providecommand \natexlab [1]{#1}%
\providecommand \enquote  [1]{``#1''}%
\providecommand \bibnamefont  [1]{#1}%
\providecommand \bibfnamefont [1]{#1}%
\providecommand \citenamefont [1]{#1}%
\providecommand \href@noop [0]{\@secondoftwo}%
\providecommand \href [0]{\begingroup \@sanitize@url \@href}%
\providecommand \@href[1]{\@@startlink{#1}\@@href}%
\providecommand \@@href[1]{\endgroup#1\@@endlink}%
\providecommand \@sanitize@url [0]{\catcode `\\12\catcode `\$12\catcode
  `\&12\catcode `\#12\catcode `\^12\catcode `\_12\catcode `\%12\relax}%
\providecommand \@@startlink[1]{}%
\providecommand \@@endlink[0]{}%
\providecommand \url  [0]{\begingroup\@sanitize@url \@url }%
\providecommand \@url [1]{\endgroup\@href {#1}{\urlprefix }}%
\providecommand \urlprefix  [0]{URL }%
\providecommand \Eprint [0]{\href }%
\providecommand \doibase [0]{http://dx.doi.org/}%
\providecommand \selectlanguage [0]{\@gobble}%
\providecommand \bibinfo  [0]{\@secondoftwo}%
\providecommand \bibfield  [0]{\@secondoftwo}%
\providecommand \translation [1]{[#1]}%
\providecommand \BibitemOpen [0]{}%
\providecommand \bibitemStop [0]{}%
\providecommand \bibitemNoStop [0]{.\EOS\space}%
\providecommand \EOS [0]{\spacefactor3000\relax}%
\providecommand \BibitemShut  [1]{\csname bibitem#1\endcsname}%
\let\auto@bib@innerbib\@empty
%</preamble>
\bibitem [{\citenamefont {Wozniak}\ and\ \citenamefont
  {Chen}(2009)}]{WozniakChen2009}%
  \BibitemOpen
  \bibfield  {author} {\bibinfo {author} {\bibfnamefont {M.~A.}\ \bibnamefont
  {Wozniak}}\ and\ \bibinfo {author} {\bibfnamefont {C.~S.}\ \bibnamefont
  {Chen}},\ }\href@noop {} {\bibfield  {journal} {\bibinfo  {journal} {Nature
  Rev. Mol. Cell Biol.}\ }\textbf {\bibinfo {volume} {10}},\ \bibinfo {pages}
  {34} (\bibinfo {year} {2009})}\BibitemShut {NoStop}%
\bibitem [{\citenamefont {Hutson}\ \emph {et~al.}(2003)\citenamefont {Hutson},
  \citenamefont {Tokutake}, \citenamefont {Chang}, \citenamefont {Bloor},
  \citenamefont {Venakides}, \citenamefont {Kiehart},\ and\ \citenamefont
  {Edwards}}]{KiehartEdwards2003}%
  \BibitemOpen
  \bibfield  {author} {\bibinfo {author} {\bibfnamefont {M.~S.}\ \bibnamefont
  {Hutson}}, \bibinfo {author} {\bibfnamefont {Y.}~\bibnamefont {Tokutake}},
  \bibinfo {author} {\bibfnamefont {M.-S.}\ \bibnamefont {Chang}}, \bibinfo
  {author} {\bibfnamefont {J.~W.}\ \bibnamefont {Bloor}}, \bibinfo {author}
  {\bibfnamefont {S.}~\bibnamefont {Venakides}}, \bibinfo {author}
  {\bibfnamefont {D.~P.}\ \bibnamefont {Kiehart}}, \ and\ \bibinfo {author}
  {\bibfnamefont {G.~S.}\ \bibnamefont {Edwards}},\ }\href@noop {} {\bibfield
  {journal} {\bibinfo  {journal} {Science}\ }\textbf {\bibinfo {volume}
  {300}},\ \bibinfo {pages} {145} (\bibinfo {year} {2003})}\BibitemShut
  {NoStop}%
\bibitem [{\citenamefont {Dembo}\ and\ \citenamefont
  {Wang}(1999)}]{DemboWang1}%
  \BibitemOpen
  \bibfield  {author} {\bibinfo {author} {\bibfnamefont {M.}~\bibnamefont
  {Dembo}}\ and\ \bibinfo {author} {\bibfnamefont {Y.-L.}\ \bibnamefont
  {Wang}},\ }\href@noop {} {\bibfield  {journal} {\bibinfo  {journal} {Biophys.
  J.}\ }\textbf {\bibinfo {volume} {76}},\ \bibinfo {pages} {2307} (\bibinfo
  {year} {1999})}\BibitemShut {NoStop}%
\bibitem [{\citenamefont {Tan}\ \emph {et~al.}(2003)\citenamefont {Tan},
  \citenamefont {Tien}, \citenamefont {Pirone}, \citenamefont {Gray},
  \citenamefont {Bhadriraju},\ and\ \citenamefont {Chen}}]{TanTienPirone}%
  \BibitemOpen
  \bibfield  {author} {\bibinfo {author} {\bibfnamefont {J.~L.}\ \bibnamefont
  {Tan}}, \bibinfo {author} {\bibfnamefont {J.}~\bibnamefont {Tien}}, \bibinfo
  {author} {\bibfnamefont {D.~M.}\ \bibnamefont {Pirone}}, \bibinfo {author}
  {\bibfnamefont {D.~S.}\ \bibnamefont {Gray}}, \bibinfo {author}
  {\bibfnamefont {K.}~\bibnamefont {Bhadriraju}}, \ and\ \bibinfo {author}
  {\bibfnamefont {C.~S.}\ \bibnamefont {Chen}},\ }\href@noop {} {\bibfield
  {journal} {\bibinfo  {journal} {Proc. Natl. Acad. Sci. USA}\ }\textbf
  {\bibinfo {volume} {100}},\ \bibinfo {pages} {1484} (\bibinfo {year}
  {2003})}\BibitemShut {NoStop}%
\bibitem [{\citenamefont {du~Roure}\ \emph {et~al.}(2005)\citenamefont
  {du~Roure}, \citenamefont {Saez}, \citenamefont {Buguin}, \citenamefont
  {Austin}, \citenamefont {Chavrier}, \citenamefont {Silberzan},\ and\
  \citenamefont {Ladoux}}]{duRoureSaezBuguin}%
  \BibitemOpen
  \bibfield  {author} {\bibinfo {author} {\bibfnamefont {O.}~\bibnamefont
  {du~Roure}}, \bibinfo {author} {\bibfnamefont {A.}~\bibnamefont {Saez}},
  \bibinfo {author} {\bibfnamefont {A.}~\bibnamefont {Buguin}}, \bibinfo
  {author} {\bibfnamefont {R.~H.}\ \bibnamefont {Austin}}, \bibinfo {author}
  {\bibfnamefont {P.}~\bibnamefont {Chavrier}}, \bibinfo {author}
  {\bibfnamefont {P.}~\bibnamefont {Silberzan}}, \ and\ \bibinfo {author}
  {\bibfnamefont {B.}~\bibnamefont {Ladoux}},\ }\href@noop {} {\bibfield
  {journal} {\bibinfo  {journal} {Proc. Natl. Acad. Sci. USA}\ }\textbf
  {\bibinfo {volume} {102}},\ \bibinfo {pages} {2390} (\bibinfo {year}
  {2005})}\BibitemShut {NoStop}%
\bibitem [{\citenamefont {Saez}\ \emph {et~al.}(2010)\citenamefont {Saez},
  \citenamefont {Anon}, \citenamefont {Ghibaudo}, \citenamefont {du~Roure},
  \citenamefont {Di~Meglio}, \citenamefont {Hersen}, \citenamefont {Silberzan},
  \citenamefont {Buguin},\ and\ \citenamefont {Ladoux}}]{SaezAnonGhibaudo}%
  \BibitemOpen
  \bibfield  {author} {\bibinfo {author} {\bibfnamefont {A.}~\bibnamefont
  {Saez}}, \bibinfo {author} {\bibfnamefont {E.}~\bibnamefont {Anon}}, \bibinfo
  {author} {\bibfnamefont {M.}~\bibnamefont {Ghibaudo}}, \bibinfo {author}
  {\bibfnamefont {O.}~\bibnamefont {du~Roure}}, \bibinfo {author}
  {\bibfnamefont {J.-M.}\ \bibnamefont {Di~Meglio}}, \bibinfo {author}
  {\bibfnamefont {P.}~\bibnamefont {Hersen}}, \bibinfo {author} {\bibfnamefont
  {P.}~\bibnamefont {Silberzan}}, \bibinfo {author} {\bibfnamefont
  {A.}~\bibnamefont {Buguin}}, \ and\ \bibinfo {author} {\bibfnamefont
  {B.}~\bibnamefont {Ladoux}},\ }\href@noop {} {\bibfield  {journal} {\bibinfo
  {journal} {J. Phys.: Condens. Matter}\ }\textbf {\bibinfo {volume} {22}},\
  \bibinfo {pages} {194119} (\bibinfo {year} {2010})}\BibitemShut {NoStop}%
\bibitem [{\citenamefont {Trepat}\ \emph {et~al.}(2009)\citenamefont {Trepat},
  \citenamefont {Wasserman}, \citenamefont {Angelini}, \citenamefont {Millet},
  \citenamefont {Weitz}, \citenamefont {Butler},\ and\ \citenamefont
  {Fredberg}}]{TrepatWasserman2009}%
  \BibitemOpen
  \bibfield  {author} {\bibinfo {author} {\bibfnamefont {X.}~\bibnamefont
  {Trepat}}, \bibinfo {author} {\bibfnamefont {M.~R.}\ \bibnamefont
  {Wasserman}}, \bibinfo {author} {\bibfnamefont {T.~E.}\ \bibnamefont
  {Angelini}}, \bibinfo {author} {\bibfnamefont {E.}~\bibnamefont {Millet}},
  \bibinfo {author} {\bibfnamefont {D.~A.}\ \bibnamefont {Weitz}}, \bibinfo
  {author} {\bibfnamefont {J.~P.}\ \bibnamefont {Butler}}, \ and\ \bibinfo
  {author} {\bibfnamefont {J.~J.}\ \bibnamefont {Fredberg}},\ }\href@noop {}
  {\bibfield  {journal} {\bibinfo  {journal} {Nature Phys.}\ }\textbf {\bibinfo
  {volume} {5}} (\bibinfo {year} {2009})}\BibitemShut {NoStop}%
\bibitem [{\citenamefont {Khalil}\ and\ \citenamefont
  {Friedl}(2010)}]{KhalilFriedl}%
  \BibitemOpen
  \bibfield  {author} {\bibinfo {author} {\bibfnamefont {A.}~\bibnamefont
  {Khalil}}\ and\ \bibinfo {author} {\bibfnamefont {P.}~\bibnamefont
  {Friedl}},\ }\href@noop {} {\bibfield  {journal} {\bibinfo  {journal}
  {Integr. Biol.}\ }\textbf {\bibinfo {volume} {2}} (\bibinfo {year}
  {2010})}\BibitemShut {NoStop}%
\bibitem [{\citenamefont {Landau}\ and\ \citenamefont
  {Lifschitz}(1986)}]{LandauLifschitz}%
  \BibitemOpen
  \bibfield  {author} {\bibinfo {author} {\bibfnamefont {L.~D.}\ \bibnamefont
  {Landau}}\ and\ \bibinfo {author} {\bibfnamefont {E.~M.}\ \bibnamefont
  {Lifschitz}},\ }\href@noop {} {\emph {\bibinfo {title} {Theory of
  Elasticity}}},\ \bibinfo {edition} {3rd}\ ed.\ (\bibinfo  {publisher}
  {Butterworth-Heinemann},\ \bibinfo {address} {Oxford},\ \bibinfo {year}
  {1986})\BibitemShut {NoStop}%
\bibitem [{\citenamefont {Murray}(2003)}]{MurrayBook2}%
  \BibitemOpen
  \bibfield  {author} {\bibinfo {author} {\bibfnamefont {J.~D.}\ \bibnamefont
  {Murray}},\ }\href@noop {} {\emph {\bibinfo {title} {Mathematical Biology
  {II}: Spatial Models and Biomedical Applications}}},\ \bibinfo {edition}
  {3rd}\ ed.\ (\bibinfo  {publisher} {Springer},\ \bibinfo {address} {New
  York},\ \bibinfo {year} {2003})\BibitemShut {NoStop}%
\bibitem [{\citenamefont {Edwards}\ and\ \citenamefont
  {Chapman}(2007)}]{EdwardsChapman}%
  \BibitemOpen
  \bibfield  {author} {\bibinfo {author} {\bibfnamefont {C.}~\bibnamefont
  {Edwards}}\ and\ \bibinfo {author} {\bibfnamefont {S.}~\bibnamefont
  {Chapman}},\ }\href@noop {} {\bibfield  {journal} {\bibinfo  {journal} {Bull.
  of Math. Biol.}\ }\textbf {\bibinfo {volume} {69}},\ \bibinfo {pages} {1927}
  (\bibinfo {year} {2007})}\BibitemShut {NoStop}%
\bibitem [{SI()}]{SI}%
  \BibitemOpen
  \href@noop {} {\emph {\bibinfo {title} {See Supplemental Material at [URL
  will be inserted by publisher].}}}\BibitemShut {Stop}%
\bibitem [{\citenamefont {Mohrdieck}\ \emph {et~al.}(2005)\citenamefont
  {Mohrdieck}, \citenamefont {Wanner}, \citenamefont {Roos}, \citenamefont
  {Roth}, \citenamefont {Sackmann}, \citenamefont {Spatz},\ and\ \citenamefont
  {Arzt}}]{SpatzSimulate}%
  \BibitemOpen
  \bibfield  {author} {\bibinfo {author} {\bibfnamefont {C.}~\bibnamefont
  {Mohrdieck}}, \bibinfo {author} {\bibfnamefont {A.}~\bibnamefont {Wanner}},
  \bibinfo {author} {\bibfnamefont {W.}~\bibnamefont {Roos}}, \bibinfo {author}
  {\bibfnamefont {A.}~\bibnamefont {Roth}}, \bibinfo {author} {\bibfnamefont
  {E.}~\bibnamefont {Sackmann}}, \bibinfo {author} {\bibfnamefont {J.~P.}\
  \bibnamefont {Spatz}}, \ and\ \bibinfo {author} {\bibfnamefont
  {E.}~\bibnamefont {Arzt}},\ }\href@noop {} {\bibfield  {journal} {\bibinfo
  {journal} {ChemPhysChem}\ }\textbf {\bibinfo {volume} {6}},\ \bibinfo {pages}
  {1492} (\bibinfo {year} {2005})}\BibitemShut {NoStop}%
\bibitem [{\citenamefont {Saez}\ \emph {et~al.}(2007)\citenamefont {Saez},
  \citenamefont {Ghibaudo}, \citenamefont {Buguin}, \citenamefont {Silberzan},\
  and\ \citenamefont {Ladoux}}]{LadouxOvalPillars}%
  \BibitemOpen
  \bibfield  {author} {\bibinfo {author} {\bibfnamefont {A.}~\bibnamefont
  {Saez}}, \bibinfo {author} {\bibfnamefont {M.}~\bibnamefont {Ghibaudo}},
  \bibinfo {author} {\bibfnamefont {A.}~\bibnamefont {Buguin}}, \bibinfo
  {author} {\bibfnamefont {P.}~\bibnamefont {Silberzan}}, \ and\ \bibinfo
  {author} {\bibfnamefont {B.}~\bibnamefont {Ladoux}},\ }\href@noop {}
  {\bibfield  {journal} {\bibinfo  {journal} {Proc. Natl. Acad. Sci. USA}\
  }\textbf {\bibinfo {volume} {104}},\ \bibinfo {pages} {8281} (\bibinfo {year}
  {2007})}\BibitemShut {NoStop}%
\end{thebibliography}
%\bibliographystyle{apsrev4-1}

%merlin.mbs apsrev4-1.bst 2010-07-25 4.21a (PWD, AO, DPC) hacked
%Control: key (0)
%Control: author (72) initials jnrlst
%Control: editor formatted (1) identically to author
%Control: production of article title (-1) disabled
%Control: page (0) single
%Control: year (1) truncated
%Control: production of eprint (0) enabled
%

\end{document}